\documentclass[nonacm]{acmart}
\setcopyright{none}

\usepackage{dcolumn}
\usepackage{bm}
\usepackage{subcaption}
\usepackage{braket}
\usepackage{mathtools}
\usepackage{hyperref}
\usepackage{xcolor}

\begin{document}

\title[Quantum Noise via Random Walk on Bloch Sphere]{Overdispersion in gate tomography: Experiments and continuous, two-scale random walk model on the Bloch sphere}

\author{Wolfgang Nowak}
  \affiliation{%
  \institution{Institute for Hydraulic and Environmental Systems (IWS/LS3) \& EXC SimTech, University of Stuttgart}
  \country{Germany}}
  \city{Stuttgart}
  \email{wolfgang.nowak@iws.uni-stuttgart.de}
\author{Tim Br\"{u}nnette}%
  \affiliation{%
  \institution{Institute for Hydraulic and Environmental Systems (IWS/LS3)}
  \country{Germany}}
  \city{Stuttgart}
  \email{tim.bruennette@iws.uni-stuttgart.de}
\author{Merel Schalkers}%
  \affiliation{%
  \institution{Delft Institute of Applied Mathematics (DIAM), Delft University of Technology}
  \country{The Netherlands}}
  \city{Delft}
  \email{M.A.Schalkers@tudelft.nl}
\author{Matthias M\"{o}ller}%
  \affiliation{%
  \institution{Delft Institute of Applied Mathematics (DIAM), Delft University of Technology}
  \country{The Netherlands}}
  \city{Delft}
  \email{M.Moller@tudelft.nl}
\renewcommand{\shortauthors}{Nowak et al.}
\date{\today}

\begin{abstract}
Noisy intermediate-scale quantum computers (NISQ) are computing hardware in their childhood, but they are showing high promise and growing quickly. They are based on so-called qubits, which are the quantum equivalents of bits. Any given qubit state results in a given probability of observing a value of zero or one in the readout process. One of the main concerns for NISQ machines is the inherent noisiness of qubits, i.e. the observable frequencies of zeros and ones do not correspond to the theoretically expected probability, as the qubit states are subject to random disturbances over time and with each additional algorithmic operation applied to them. Models to describe the influence of this noise exist.

In this study, we conduct extensive experiments on quantum noise. Based on our data, we show that existing noise models lack important aspects. Specifically, they fail to properly capture the aggregation of noise effects over time (or over an algorithm's runtime), and they are underdispersed. With underdispersion, we refer to the fact that observable frequencies scatter much more between repeated experiments than what the standard assumptions of the binomial distribution would allow for. Based on these shortcomings, we develop an extended noise model for the probability distribution of observable frequencies as a function of the number of gate operations. The model roots back to a known continuous random walk on the (Bloch) sphere, where the angular diffusion coefficient can be used to characterize the standard noisiness of gate operations. Here, we superimpose a second random walk at the scale of multiple readouts to account for overdispersion. Further, our model has known, explicit components for noise during state preparation and measurement (SPAM). The interaction of these two random walks predicts theoretical, runtime-dependent bounds for probabilities. Overall, it is a three-parameter distributional model that fits the data much better than the corresponding one-scale model (without overdispersion), and we demonstrate the better fit and the plausibility of the predicted bounds via Bayesian data-model analysis. 
\end{abstract}


\maketitle

% ===========================================================+
% ===========================================================+
\section{Introduction\label{sec:Intro}}
% ===========================================================+
% ===========================================================+

% qubits versus bits
Quantum computing is an emerging computing technology that has the potential to revolutionize the way we solve computational problems in the future. The potential power of quantum computing stems from the fact that information is stored in so-called quantum bits (qubits) which, unlike their binary counterparts can hold a superposition of the two orthonormal basis states $\ket{0}$ and $\ket{1}$. However, qubit states as such are not observable. Instead, when a qubit is measured, it collapses to either the zero or one state with a \emph{probability} that relates to the superimposed state (see Section~\ref{sec: Methods}).

% multi-qubit and coolness
The above-described concept generalizes to quantum registers. An $n$-qubit register can hold a superposition of all $2^n$ bitstrings $\ket{00\dots00},\, \ket{00\dots01},\,\dots,\,\ket{11\dots11}$ which makes it possible to perform operations on an exponential amount of input data simultaneously. Unlike classical computing, this quantum parallelism does neither require an exponential overhead in time nor the replication of hardware in space to enable parallel execution. It is a result of the qubits' capability to hold multiple states simultaneously.

% readout is the bottleneck
Quantum computing cannot surpass the no-free-lunch theorem, which manifests itself in the retrieval of information from the output state. As states cannot be observed directly, quantum algorithms need to be executed multiple times to obtain an approximation of the probabilities from which the most probable `outcome(s)' of the quantum algorithm can be deduced.

% NISQ is noisy!
Today's noisy intermediate-scale quantum (NISQ) computers are not yet ready for solving meaningful real-world problems due to the immaturity of the quantum hardware. State-of-the-art quantum computers from IBM provide up to 433 qubits, which enables the superposition of 2.2e130 states and should bring the solution of real-world problems into reach \cite{IBM}. However, qubits are very susceptible to noise and feature extremely short decoherence times. This noise limits the complexity of quantum algorithms to perform a few basic single- and two-qubit operations (to be explained below) before the output state degenerates to meaningless noise.

% noise depends on the machine.
The noise that occurs in practical quantum computation is manifold and can be more or less pronounced for different qubit technologies such as superconducting qubits, trapped ion qubits, photonic qubits, neutral-atom qubits, and semiconductor spin qubits \cite{Wilen2021, Monz2010, Sharma2021, Wintersperger2023, Shehata2023}. Without going too much into the technical details one can differentiate at least between two types of errors:
\begin{itemize}
    \item \emph{systematic errors} such as the mistuned microwave pulse leading to a systematic under-rotation, over-rotation, or rotation around an imprecise axis
    
    \item \emph{spurious cross-talk} between neighboring qubits when applying an operation to one or more of them or measuring one or more qubit
\end{itemize}

However, the standard noise models implemented in today's quantum computer simulators such as depolarizing channel models \cite{nielsen2010quantum} are typically very simplistic and not capable of modeling the complex noise behavior.

Thus, useful methods for quantum characterization, verification, and validation (QCVV) are highly relevant.

% SOTA: RB
A commonly used and scalable method to partially characterize the noise associated with quantum gates is randomized benchmarking (RB) \cite{Helsen2022}. 
RB was originally proposed in 2005 \cite{Emerson2005} with different versions and flavors coming out ever since \cite{Eisert2020}.  In RB, the average gate fidelities of a given NISQ machine (on average over many types and repetitions of gates) are found by performing a randomized series of mutually canceling operations of varying lengths. Upon measurement, one theoretically expects to always find the initial (zero) state but does not in practice. The exponential decay rate explaining these imperfect measurements is the core error metric. One known downside of this technique is that it makes the assumption that noise is time-independent \cite{Wallman_2014}.

% SOTA: GST
Tomographic methods aim to characterize the quantum operations including noise in full detail. While some of the methods focus on specific parts of the quantum computations (such as state tomography, process tomography, or measurement tomography), the widely used gate set tomography (GST) aims to reconstruct the entire pipeline, including state preparation and measurement errors (SPAM) \cite{Nielsen2021gatesettomography}. While the comprehensiveness of this approach has its appeal, the level of detail requires large, sometimes prohibitively large amounts of experimental data.

% What will be new with us?
Our proposed method, data (see Section~\ref{subsec: data}) and model (see Section~\ref{subsec: TwoLevelRW}) are akin to RB in that they use lumped error characteristics, with the twist that these are happening at multiple levels. We model the qubits' movement on the Bloch sphere with errors occurring on an individual as well as a pool level while also considering finite sample effects. The two levels roughly represent random errors that affect individual qubits only and random effects, possibly caused by environmental factors, affecting batches of qubits identically. This combination helps to explain overdispersion, an effect noticed in \cite{Nielsen_2021_Nestederror} and \cite{ostrove2022Overdispersion}, in an easily accessible manner while avoiding a generic catch-all error. Further, the model could easily allow an extension to incorporate non-Markovian error types \cite{blume-kohout2022markovian}.
So far, we focus on the characterization of one particular basis gate, however, an application to random gate sequences, as commonly used in RB\cite{Emerson2007}, is possible.

% Research questions:
To explore our idea, we pose the following research questions:
    \begin{itemize}
        \item Can the experimental data confirm that we need two levels?
        
        \item Our hypothesized model will predict growth and then decay of overdispersion over algorithm runtime, and even provides theoretical bounds for the overdispersion at any given algorithm runtime. Can we observe evidence of this?
    \end{itemize}

% ===========================================================+
% ===========================================================+
\section{Methods\label{sec: Methods}}
% ===========================================================+
% ===========================================================+

% ===========================================================+
\subsection{Bloch sphere representation for a single qubit\label{subsec: BlochSphere}}
% ===========================================================+

% superposition
A qubit is the quantum-computing version of a bit. Its state is generally denoted as $\left|\Psi\right>$, defined by:

\begin{eqnarray}
\label{eq:QubitStatePsi}
    \left|\Psi\right> = \gamma_0 \left|0\right> + \gamma_1 \left|1\right>\, , \quad \gamma_0, \gamma_1 \in \mathbb{C} , \quad
    \left|0\right> = \begin{bmatrix}1 \\  0 \end{bmatrix} \, , \;
    \left|1\right> = \begin{bmatrix}0 \\  1 \end{bmatrix} \, ,
\end{eqnarray}

where the complex numbers $\gamma_0$ and $\gamma_1$ are sometimes called probability amplitudes. Furthermore, $\left|0\right>$ and $\left|1\right>$ are orthonormal basis states forming the computational basis. Eq.~\eqref{eq:QubitStatePsi} represents a superposition of the two basis states.

% measurement probabilities
Qubit states as such are not observable. When a qubit is measured, it collapses to either $\left|0\right>$ or $\left|1\right>$ (here simply called 0 or 1) with probabilities $P$ given by:

\begin{eqnarray} \label{eq:PfromAB}
 P[0] =|\gamma_0|^2 \; , \quad  P[1] = |\gamma_1 |^2.
\end{eqnarray}

As the events to measure 0 or 1 are mutually exclusive and collectively exhaustive, the corresponding probabilities sum up to unity:

\begin{equation} \label{eq:TotalProbability}
    |\gamma_0|^2 + |\gamma_1|^2 = P[0]+P[1] = 1.
\end{equation}

% representation on the Bloch sphere
From the total of four degrees of freedom contained in the two complex numbers $\gamma_0$ and $\gamma_1$, the constraint in Eq.~\eqref{eq:TotalProbability} removes one. Another degree of freedom has no physically observable consequences, and so a qubit has two relevant degrees of freedom \cite{nielsen2010quantum}. Due to Eq.~\eqref{eq:TotalProbability}, these form a unit sphere, called the Bloch sphere \cite{nielsen2010quantum}, which can be thought of as a sphere in $\mathbb{R}^3$.

% Geology of the Bloch sphere
The Bloch sphere is like a globe, where the north pole represents the zero-state $\left|0\right>$ and the south pole represents the one-state $\left|1\right>$. The so-called colatitude is given by the angle $\theta \in [0,\pi]$ with zero at the north pole (in contrast to regular latitude). For $\theta=0$ at the north pole, a measurement returns the zero-state with probability $P[0]=1$ and similarly at the south pole with $\theta=\pi$. The equator at $\theta=\pi/2$ corresponds to $P[1]=P[0]=1/2$. The longitude is given by the angle $\varphi \in [0,2\pi]$, which does not influence the probabilities at all, but is still relevant in quantum computing as it changes the readout probabilities after appropriate rotations. 

% Angles on the Bloch sphere
Since the Bloch sphere is a powerful visualization and a good basis for constructing noise models, it is worthwhile to note the transformations between the probability amplitudes $\gamma_0$ and $\gamma_1$ and the angles $\theta$ and $\varphi$:

\begin{equation}
 \label{eq:ABfromAngles}
 \gamma_0 = \cos\left(\frac{\theta}{2}\right) \; , \quad 
 \gamma_1  = e^{i\varphi} \sin\left(\frac{\theta}{2}\right).
\end{equation} 

As the longitude $\varphi$ does not influence the probabilities $P[0]$ and $P[1]$, we look specifically at the relation between $\theta$ and $P=P[0]$:

\begin{eqnarray} \label{eq:PfromTheta}
 P=P[0] &=& |\gamma_0|^2 = \cos^2\left(\frac{\theta}{2}\right) 
        = \frac{1}{2}+\frac{1}{2} \cos\left(\theta\right) \nonumber \\
        \label{eq:ThetaFromP}
 \Rightarrow \quad \theta &=& \arccos \left( 2P-1\right).
\end{eqnarray}

For transforming probability density functions between $\theta$ and $P$, the Jacobian of this relation will be useful:

\begin{eqnarray} 
\label{eq:Jacobian}
 \frac{\partial \theta}{\partial P} &=& - \frac{1}{\sqrt{P-P^2}} \; .
\end{eqnarray}

The relation is monotonic, which is an important prerequisite in transformations of random variables \cite[e.g.,][]{casella2021statistical}.

% ===========================================================+
\subsection{Data\label{subsec: data}}
% ===========================================================+

% -----------------------------------------------------------+
\subsubsection{Data creation}
% -----------------------------------------------------------+

% canceling gates
To build and test our extended noise model, we first acquire the necessary data by running experiments on open-access quantum computers. The experiments are set up such that, in noise-free theory, we know we should always measure the zero-state with probability 1. 
To obtain observed frequencies (as the probabilities cannot be directly observed), we repeat each algorithm 8192 times, which is the upper limit on the freely available machine we use. We then repeat this for multiple circuits with varying lengths (i.e. varying numbers of gate operations), always ensuring that the expected theoretical outcome remains the zero state.

% numbers of gates
We collect the resulting data and plot the observed frequency of measuring the zero-state as a function of the number of gates applied to the qubit. In this paper, we consider each such gate applied as a time step, therefore the axis representing the number of gates applied can be interpreted as a time axis.

% choice of gate
In the data for our model, we make use of the $\sqrt{x}$-gate, one of IBM's native gates. In each considered circuit, the $\sqrt{x}$-gate is applied a number of times that is a multiple of four, ensuring that the total effect on the qubits should vanish~\footnote{Note that $\sqrt{x} = \frac{1}{2} \begin{bmatrix} 1+i & i-1 \\ 1-i & 1+i \end{bmatrix} $, therefore $\sqrt{x} \sqrt{x} \sqrt{x} \sqrt{x} = I$}. The circuits are compiled using optimization level-0, which guarantees that the gates are actually performed on the physical qubit(s) (and not optimized away as they cancel each other out).

\begin{figure}
     \centering
     \begin{subfigure}[b]{0.48\textwidth}
         \centering
         \includegraphics[width=\textwidth]{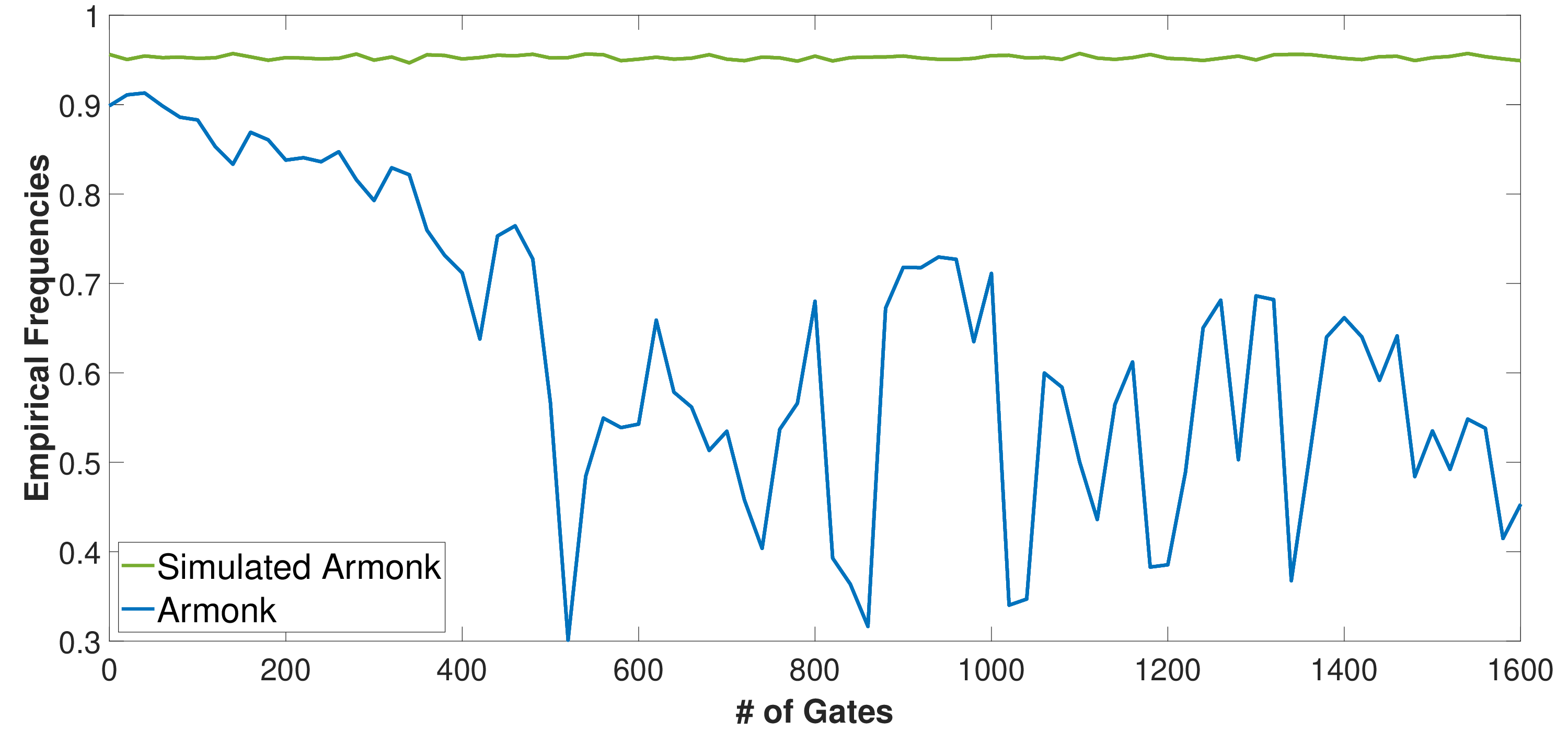}
         \caption{Results from the simulated and real Armonk device. For the real device, each data point was produced at a different wall clock time.}
         \label{fig:singlerunspread}
         \Description{While no significant error accumulates for the simulated Armonk machine, the real machine shows some noisy behavior. }
     \end{subfigure}
     \hfill
     \begin{subfigure}[b]{0.48\textwidth}
         \centering
         \includegraphics[width=\textwidth]{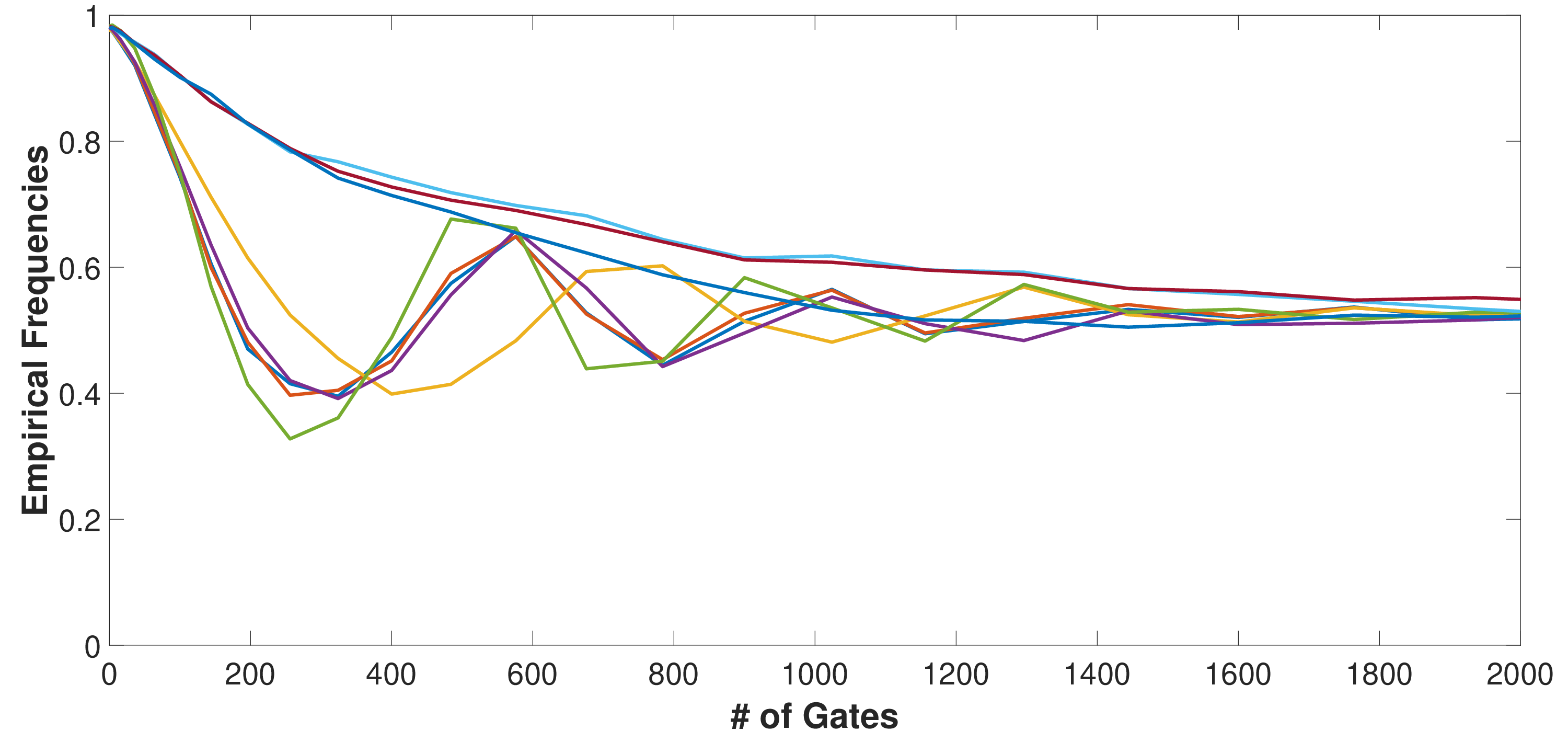}
         \caption{Multiple runs of repeated $\sqrt{x}$ applications. Each curve corresponds to a single queue, thus effectively a single wallclock time.}
         \label{fig:multirun}
         \Description{If the experiments for the real machine are scheduled in one queue, the error is more structured, sometimes approaching the equilibrium from one side and sometimes in a clearly coherent oscillatory behavior.}
     \end{subfigure}
        \caption{Plots showing the different noise behaviors in three cases. The simulated device, the real machine with mixed wallclock times and the real machine with (multiple) single wallclock times. }
\end{figure}

% preview of data
In Figure \ref{fig:singlerunspread}, we plot the frequencies of measuring the correct (here: zero) state in the case that the experiments for the different data points are run at different wallclock times. In Figure \ref{fig:multirun}, we plot the frequencies of measuring the correct state in the case that the experiments for the different data points are run at the same wallclock time. Here, ``different time" means through different jobs submitted to the system's waiting queue that have been processed in intervals of, on average, seven minutes between the individual jobs; ``same time" means submitted as a bulk job without any waiting time in between. Regardless of this difference, one can see how the degree of dispersion (i.e. the scatter of the lines around the hypothetical mean line) increases at first, until somewhere before 500 gate operations in the specific case, and then decreases again.

% -----------------------------------------------------------+
\subsubsection{Data availability}
% -----------------------------------------------------------+

% detailed settings
The full settings used for 1-qubit IBM Armonk machine \cite{IBM} can be found in Appendix \ref{app:ArmonkSettings}.
% data 
The data produced is available (\emph{link to GitLab repository will be released after publication}) \href{www.github.com/MerelSchalkers/QubitNoiseModel}{online}. The data file contains the resulting frequency of measuring the zero-state, the amount of $\sqrt{x}$-gates applied, and the timestamp of when the experiment was run on the IBM Armonk machine.

% ===========================================================+
\subsection{Relevant levels of randomness and probability distributions\label{subsec: LevelsOfRandomness}}
% ===========================================================+

In the following derivations, four distinct levels of randomness will appear. To avoid confusion, we declare them upfront:

\begin{itemize}
    \item \textbf{Bernoulli level}: Consider the state $\left|\Psi\right>$ (here: specifically $\theta$) of a qubit to be fixed. For an individual measurement $(n=1)$, we know the theoretical probability $P[0]$ via Eq.~\eqref{eq:PfromTheta}. But when conducted we see either a zero or one, with no meaningful way to define a (frequentist) probability in a single experiment due to $n=1$.
    
    \item \textbf{Binomial level}: Again consider a qubit in fixed state with theoretically known $P$. After $n>1$ repetitions, one can observe a meaningful frequency of the binary outcomes $\{0,1\}$. For $n\rightarrow \infty$, the observed frequency converges to the Bernoulli-level probability. For distinguished notation, we use $\bar{P}=P$ for probabilities at the Binomial level and $\bar{F}$ as observed frequencies. The discrepancy between $\bar{P}$ and $\bar{F}$ is described by the well-known Binomial distribution \cite{forbes2011statistical}. 
    
    \item \textbf{Pool level}: Consider a total of $m$ repetitions of the Binomial level (with $n$ repetitions each). These $m$ "outer" repetitions form a so-called pool of observed frequencies $\bar{F}_q; q=1\ldots m$. Each of the $m$ repetitions can have a different qubit state with corresponding theoretical (Binomial-level) probability $\bar{P}_q, q=1\ldots m$. When analyzing the pooled data altogether, one observes a pooled frequency $\check{F}$, which converges for $n,m\rightarrow\infty$ towards the theoretical mean $\check{P}$ of the Binomial-level probabilities $\bar{P}_q$. However, the observed $\bar{F}_q$ will not only scatter around $\check{P}$ due to the classical Binomial law but also due to differences in the respective values $\bar{P}_q$ and due to finite $m$; therefore, the Binomial distribution does not apply to such pooled data.
    
    \item \textbf{Parametric level}: The levels so far are all classical frequentist uncertainties. Probability distribution models fitted to the observed data are typically governed by parameters. Parametric uncertainty refers to the fact that these distributional parameters can only be inferred with finite precision at finite $m,n$. The corresponding parametric uncertainty is often quantified via Bayesian inference \cite{box2011bayesian}.
\end{itemize}

Relevant, well-known probability distributions used or referred to in the upcoming derivations are:

\begin{itemize}
    \item The Binomial distribution describes the observable frequency under $n$ independent repetitions of a binary-outcome (e.g., ${0,1}$) random experiment with constant theoretical probability $P$: $\bar{F}\sim Bino(n,P)$, with parameters $n,P$.
    
    \item The Uniform distribution $X\sim \mathcal{U}(0,1)$ assigns identical probability density to all values of a continuous random variable $X$ in the interval $[0,1]$ \cite{forbes2011statistical}.

    \item Gilbert's sine distribution $X\sim Sine() = \sin(2x) \, , \; \forall x \in [0, \pi/2]$ assigns a sinus-shaped probability density \cite{edwards_gilberts_2000}. It can be scaled to any interval $[L,U]$ with width $W=U-L$ by $Sine(L,U)=W/2\pi \cdot \sin (2\pi/W \cdot 2(x+L))$.
    
\end{itemize}

% ===========================================================+
\subsection{Two-level random walk on the Bloch sphere\label{subsec: TwoLevelRW}}
% ===========================================================+

% idea of RW on the sphere
As a possible noise model, we see the noise of qubit states as a random walk on the Bloch sphere: Imagine a qubit initialized at the north pole. At later times, the position on the sphere will not necessarily be identical to the starting position; instead, the position on the sphere is randomly moving over time, e.g., due to noisy gate operations.

% imagine the second-level RW
The new model to account for overdispersion proposed here is a two-level random walk on the Bloch sphere. The first level represents classical noise ideas as just described, while the second level, the pool level, is responsible for the overdispersion. Graphically, one can imagine the pool-level random walk as an additional, joint component in the random walks of many random walkers, specifically a joint motion of their center of mass on the surface of the sphere. 

% announcing the reduced length
The average position over many random positions of pure states \emph{on} the Bloch sphere due to the first level of the random walk actually lies \emph{within} the sphere. That means it has a shortened radius smaller than one. We will exploit this fact later to summarize the relevant statistics of the first random walk and combine it with the second random walk.

%-------------------------------------------------------------
\subsubsection{Known results for $\theta({t})$ in single-level random walks\label{sssec:KnownResults}}
%-------------------------------------------------------------

% Intro
In quantum computing, a probability distribution over states is called a mixed or \emph{decoherent} state, while a deterministic state according to Eq.~\eqref{eq:QubitStatePsi} is called a \emph{coherent} state \cite{nielsen2010quantum}. At the Binomial level (see Section~\ref{subsec: LevelsOfRandomness}), we define a theoretical probability $\bar{P}$. This $\bar{P}$ is the expected value of the noise-affected Bernoulli-level probabilities $P$ (see Section~\ref{subsec: LevelsOfRandomness}) over the probability distribution of qubit states $p(\theta,\varphi)$: $\bar{P}=E_{\theta, \varphi} [|\gamma_0|^2]$. Such a distribution on the sphere is in fact equivalent to a mixed state, and could as well be represented by a shortened Bloch vector, or by density matrices \cite{nielsen2010quantum}, i.e. as a decoherent state. However, as only $\theta$ is the probability-relevant angle, we can afford to simplify our derivations.

%Random walk model
As a starting point for our noise model, assume a random walk on the Bloch sphere, here written as time-discrete:

\begin{equation}\label{eq:RandomWalkOnSphere}
    \left| \Psi \right>_{t+1}
    =  \left| \Psi \right>_{t} 
    + \Delta_{\left| \Psi \right>,t} \, .
\end{equation}

Further, assume that the increments $\Delta_{\left| \Psi \right>,t}$ are independent and identically distributed for any given time lag $\Delta t = t_i-t_j$. Due to the independence property, the random series $\left| \Psi \right>_{t}$ is a Markov process \cite[e.g.,][]{karlin2014first}. If, additionally, we assume that these increments are Gaussian distributed with zero mean and fixed variance, and we define these increments as angular increments, we arrive at Fickian diffusion on the Bloch sphere. Written in terms of angles $\theta,\varphi$, it is governed by:

\begin{equation}
\frac{\partial p(\theta,\varphi;t)}{\partial t} = D_n \nabla p(\theta,\varphi;t), 
\end{equation}

where $p(\theta,\varphi;t)$ is the joint probability density function (PDF) of $\theta$ and $\varphi$ as a function of time $t$,  $D_n$ is the rotational diffusion coefficient in $\text{rad}^2/\text{s}$.

% assumptions towards an analytical solution
We restrict ourselves to initial conditions of $\theta=0$. Also, as we do not consider systematic actions on $\varphi$ that would make $\varphi$ relevant to probabilities $P$, we can afford to neglect $\varphi$ as of now. A small discussion on the limitation due to this omission can be found in section \ref{sec:conclusion}. So let $p(\theta;t)$ be the PDF for the colatitude $\theta$ as a function of time $t$. For the case of constant $D_n$ and the initial condition $p(\theta;t=0)=\delta(\theta)$, expansion in spherical harmonics provides the following analytical solution, as derived in \cite{saragosti_modeling_2012}:

\begin{eqnarray}\label{eq:thetaRQonSphere}
    p(\theta;t) &=& \sum_{k=0}^{\infty} \frac{2k+1}{2} e^{-D_n k (k+1) t} \cdot L_k(\cos\theta) \sin\theta \quad \forall t>0,
\end{eqnarray}

where $L_k$ is the Legendre polynomial of order $k$ \cite{abramowitz1968handbook}. In practice, the infinite sum has to be truncated and is known to converge only slowly for small $t$. However, with fast and accurate algorithms for evaluating Legendre polynomials, truncation at $K=1000$ or higher is unproblematic.

%-------------------------------------------------------------
\subsubsection{Mapping onto $P(t)$\label{sssec:MappingToP} (Bernoulli level)}
%-------------------------------------------------------------

% PDF of probabilities
While the above analytical solution is a known result, we are interested in transforming it into the PDF for the Bernoulli-level measurement probability $P=P[0]$. For clarity, we will now write $p_{\theta}(\theta)$ and $p_P(P)$, where one must realize that $p_P(P)$ is a probability density that describes what theoretical probabilities $P$ one may encounter. Using standard rules for PDFs after change of variables \cite{casella2021statistical}, we have:

\begin{equation}
    p_P(P;t) =
    p_{\theta}\bigg(\theta(P);t\bigg)
    \cdot
    \left| \frac{\partial\theta(P)}{\partial P}   \right|.
\end{equation}

Recalling the transformation in Eq.~\eqref{eq:ThetaFromP} and its Jacobian in Eq.~\eqref{eq:Jacobian}, inserting into Eq.~\eqref{eq:thetaRQonSphere} yields

\begin{equation} \label{eq:pPPt}
    p_P(P;t) = \sum_{k=0}^{\infty} (2k+1) e^{-D_n k (k+1) t} L_k(2P-1), 
\end{equation}

since

\begin{eqnarray}
    \cos \theta = 2P-1 \; , \quad 
    \sin \theta &=& 2\sqrt{P-P^2}.
\end{eqnarray}

%-------------------------------------------------------------
\subsubsection{Limit cases and information entropy\label{sssec:LimitCases}}
%-------------------------------------------------------------

For a better understanding, it is worthwhile to look at certain limit cases of Eq.~\eqref{eq:pPPt}:

\begin{eqnarray}\label{eq:LimitCasesPP}
    \lim_{t\rightarrow \infty} p_P(P;t) &=& L_0(2P-1) = 1 \, , \; \forall P\in [0,1] \\
    \label{eq:LimitCasesPP_P0}
    p_P(P=0;t) &=&
        \sum_{k=0}^{\infty} (-1)^k(2k+1) e^{-D_n k (k+1) t} \\
    \label{eq:LimitCasesPP_P1}
    p_P(P=1;t) &=&
        \sum_{k=0}^{\infty} \phantom{(-1)^k}(2k+1) e^{-D_n k (k+1) t} , 
\end{eqnarray}

which is due to  $L_k(-1)=(-1)^k$ and $L_k(1)=1$, respectively. Looking at the first limit, we see that qubit states approach a \emph{uniform} distribution over their measurement probabilities at late times, the distribution with maximal information entropy, i.e. the least informative one. At the same time, Eq.~\ref{eq:LimitCasesPP_P0} approaches its $t\rightarrow \infty$ limit of $1$ strictly from above, and Eq.~\ref{eq:LimitCasesPP_P1} approaches the same $t\rightarrow \infty$ limit, but strictly from below and starting from zero.

% max entropy in probabilities
Upon reflection, Eq.~\ref{eq:LimitCasesPP} is intuitive. Due to symmetry, the theoretical average of the uniform distribution is $0.5$. Therefore, the observable frequencies over many measurements (each one with a different, random probability from the uniform distribution) converge to the intuitive uninformative 50:50 coin toss between $\left|0\right>$ and $\left|1\right>$. Such types of averages over \ref{eq:pPPt} will be treated in more detail in the upcoming section. But, as a noteworthy fact, the random walk on the Bloch sphere installs the maximum-entropy property (here: \emph{uniform} distribution) in terms of measurable probabilities, and not in terms of $\theta$:

\begin{eqnarray}
    \lim_{t\rightarrow\infty} p_{\theta}(\theta;t)= \frac{1}{2}\sin(\theta)\, ,
\end{eqnarray}

which ends up as the $Sine()$ distribution on the $[0,\pi/2]$ interval. A uniform distribution on the \emph{surface} of the Bloch sphere is not the same as a uniform distribution on the \emph{angles}.

%-------------------------------------------------------------
\subsubsection{Transfer to theoretical probabilities $\bar{P}(t)$ (Binomial level)\label{sssec:TransferTheoProb}}
%-------------------------------------------------------------

% from individual probability to Binomial P
Eq.~\eqref{eq:pPPt} is a distribution over the individual probabilities of single ($n=1$) measurements at the Bernoulli level. Accordingly, each individual measurement will result in an observed zero-state or one-state with individual random probability $P$ and $1-P$, respectively. Therefore, the probability distribution as in Eq.~\eqref{eq:pPPt} will never be observable. Instead, when repeating the measurement process $n$ times, with $n\rightarrow \infty$, one will empirically observe a probability $\bar{P}(t) = E[P(t)]$ at the Binomial level. The fact that the empirically visible probability is the expected value $E[P(t)]$ follows from the properties of the Poisson-Binomial distribution \citep{wang_PoiBin_1993}.

% taking moments
Hence, we are now interested in the expected value $\mu_P=E[P(t)]$ and the general $\ell$th moments $\mu_{P,\ell}(t)=E[P^{\ell}(t)]$ of Eq.~\eqref{eq:pPPt}:

\begin{eqnarray}
    \mu_{P,\ell}(t)&=&E[P^{\ell}(t)] = \int_{0}^{1} p_P(P;t) P^{\ell}\,\, \text{d}P \nonumber \\
    &=& \int_{0}^{1} \sum_{k=0}^{\infty} (2k+1) e^{-D_n k (k+1) t} P^{\ell} L_k(2P-1) \,\, \text{d}P  \nonumber  \\
     &=& \sum_{k=0}^{\infty} (2k+1) e^{-D_n k (k+1) t} \int_{0}^{1}  P^{\ell} \mathcal{L}_k(P) \,\, \text{d}P\, \nonumber \\
\end{eqnarray}

with the \emph{shifted} Legendre polynomials $\mathcal{L}_k=L_k(2x-1)$, whose orthogonality property $\int_{0}^{1} \mathcal{L}_{\ell}(x) \mathcal{L}_k(x) \, \text{d}x=(2k+1)^{-1} \delta_{\ell k}$ \cite{abramowitz1968handbook} we exploit in the following.
% zeroth moment
Specifically for $\ell=0$ we have $P^{0}=\mathcal{L}_0(P)$, and hence
\begin{eqnarray}
    \mu_{P,0}(t)
    &=& \sum_{k=0}^{\infty} (2k+1) e^{-D_n k (k+1) t} \; \cdot \underbrace{\int_{0}^{1}  \mathcal{L}_0(P) \mathcal{L}_k(P) \,\, \text{d}P\,}_{(2k+1)^{-1} \delta_{k0}} 
    = 1 \quad \forall t\, ,
\end{eqnarray}
which is a requirement for valid PDFs.

% first moment
The first interesting case is for $\ell=1$, giving the probability $\bar{P}(t)$. Here, we use $P^1=\frac{1}{2}\mathcal{L}_0(P)+\frac{1}{2}\mathcal{L}_1(P)$:

\begin{eqnarray} \label{eq:Pt_exponential}
    \bar{P}(t)
    &=& \mu_{P,1}(t)
    = \sum_{k=0}^{\infty} (2k+1) e^{-D_n k (k+1) t} \cdot \underbrace{\int_{0}^{1}  \left(\frac{1}{2}\mathcal{L}_0(P)+\frac{1}{2}\mathcal{L}_1(P)\right) \mathcal{L}_k(P) \,\, \text{d}P\,}_{
    \frac{1}{2}(2k+1)^{-1} \delta_{k0} + 
    \frac{1}{2}(2k+1)^{-1} \delta_{k1} }
    \nonumber \\
    &=& \underbrace{\frac{1}{2}              }_{k=0}
     + \underbrace{\frac{1}{2} e^{-2 D_n  t}}_{k=1} \, .
\end{eqnarray}

This is a beautiful result; for qubit states subject to a random walk on the Bloch sphere with isotropic, constant angular diffusion coefficient $D_n$ after initialization to the north pole, the measurements asymptotically approach the indifferent 50:50 probability with the first-order rate $2D_n$, allowing to describe the development by

\begin{eqnarray}
    \frac{\partial \bar{P}(t)}{\partial t}
    &=& -2D_n \left(\bar{P}-\frac{1}{2}\right) \, .
\end{eqnarray}

One should remain clear that this does not imply that qubit states as such tend to the equator. Instead, the ensemble average over all possible noise-affected states tends towards the uninformative 50:50 probability, when the qubit states are uniformly distributed over the Bloch surface, i.e. an entirely decoherent state.

% second raw moment
The second moment will play a role in overdispersed data at the pool level. Hence, we set $\ell=2$, use $P^2=\frac{1}{3}\mathcal{L}_0(P)+\frac{1}{2}\mathcal{L}_1(P)+\frac{1}{6}\mathcal{L}_2(P)$ and repeat the procedure:

\begin{eqnarray}
    \mu_{P,2}(t)
    &=& \sum_{k=0}^{\infty} (2k+1) e^{-D_n k (k+1) t} \cdot \int_{0}^{1}  \left(
    \frac{1}{3}\mathcal{L}_0+\frac{1}{2}\mathcal{L}_1+\frac{1}{6}\mathcal{L}_2
    \right) \mathcal{L}_k \,\, \text{d}P\, 
        \nonumber \\
     &=& \underbrace{\frac{1}{3}              }_{k=0}
     + \underbrace{\frac{1}{2} e^{-2 D_n  t}}_{k=1}
     + \underbrace{\frac{1}{6} e^{-6 D_n  t}}_{k=2} \, .
\end{eqnarray}

% raw moment to variance
By transformation from raw moments $\mu_{\ell}$ to central moments $\mu_{\ell c}$ \cite[e.g.,][]{papoulis2002probability}, we obtain the variance $\sigma^2_{P}(t)$ for the Bernoulli-type probabilities $P$:

\begin{eqnarray}
    \sigma^2_{P} &=& \mu_{P,2c} = \mu_{P,2} - \mu_{P,1}^2 = \frac{1}{3}  + \frac{1}{2} e^{-2 D_n  t}
      + \frac{1}{6} e^{-6 D_n  t} - \left( \frac{1}{2} + \frac{1}{2} e^{-2 D_n  t}\right)^2 \nonumber \\
  &=&   \frac{1}{12}  -\frac{1}{4} e^{-4 D_n  t}
      + \frac{1}{6} e^{-6 D_n  t}  \label{eq:overdispersion}
\end{eqnarray}

% Variance special cases
Special cases are:

\begin{eqnarray}
  \sigma^2_P(t=0) &=& 0 \\
  \lim_{t\rightarrow\infty} \sigma^2_P(t) &=& \frac{1}{12}\, \, ,
\end{eqnarray}

where the former is due to the deterministic initial condition, and the latter is known to be the variance of the uniform distribution \cite{forbes2011statistical}, the limit distribution as discussed in Section \ref{sssec:LimitCases}.

%-------------------------------------------------------------
\subsubsection{Pool-level randomness\label{sssec:PoolLevel}}
%-------------------------------------------------------------

% Definition of pool
For extending the above results to overdispersed data, we now invoke the pool level with $m$ repetitions of a Binomial-level experiment. Each Binomial-level experiment with $n$ measurements at a given time $t$ has a different Binomial-level probability $\bar{P}_q(t), q=1\ldots m$. 
We now assume that the random walk on the Bloch sphere has two components: (1) an i.i.d. noise component as discussed above, which works individually on all Bernoulli trials; (2) a common Binomial-level component for all $n$ random walkers of a size-$n$ Binomial experiment that introduces randomness into a size-$m$ pool of Binomial experiments, causing the probabilities $\bar{P}_q(t)$ within the pool to vary.

% Reduced-Length argument for the lower level
For all linear relations $y=f(x)$, the simplification $E[y]=f(E[x])$ can be used. We can use this simplification due to the linear character of qubit gates that represent rotations (and hence random moves) on the Bloch sphere. Hence, we can directly work with the \emph{expected} position from the Binomial-level random walk on the Bloch sphere to construct our pool-level random walk. The expected position, due to symmetry, stays constant $\theta(t)=0$ at the north pole for all times but simply has a reduced \emph{length} $R$ of the Bloch vector given by

\begin{eqnarray}
    R(t) = 1-2\bar{P}(t)
    &=& e^{-2 D_n  t} \, ,
    \label{eq:reducedLength}
\end{eqnarray}

which follows from Eq.~\eqref{eq:Pt_exponential}. We use this reduced length to represent the statistics of the first level of random walk at the Binomial level. Then, we use Eq.~\eqref{eq:pPPt} anew for representing the second level, just working on the reduced-length vector. Thus, we construct the overdispersed probability distribution of $\bar{P}_q(t)$ by using Eq.~\eqref{eq:thetaRQonSphere} together with the reduced length from Eq.~\eqref{eq:reducedLength}:

\begin{equation}
    \bar{P}_q(\theta_q;t) = \frac{1}{2} + R(t) \cdot \frac{1}{2} \cos(\theta_q) \; ,
    \label{eq:overdisp_Pbar}
\end{equation}

where $\theta_q$ is the random position of the joint center of mass of all $n$ random walkers in the size-$n$ Binomial experiment number $q$. While $\theta_q$ played no role before and always stayed at the north pole, as of now, it is subject to its own random walk that also follows Eq.~\eqref{eq:RandomWalkOnSphere}.

% Deriving overdispersed distribution
Next, we follow all derivation steps from Eq.~\eqref{eq:PfromTheta} and \eqref{eq:Jacobian} to Eq.~\eqref{eq:pPPt}, but with Eq.~\eqref{eq:overdisp_Pbar} as the starting point to obtain:

\begin{equation}
\label{eq:pPPtreducedLength}
    p_{\bar{P}_q}(\bar{P}_q;t) = p_{\theta_q}(\theta_q;t) |\frac{\partial \theta_q}{\partial \bar{P}_q}| 
 = \frac{1}{R(t)} \sum_{k=0}^{\infty} (2k+1) e^{-D_q k (k+1) t}  L_k \left(\frac{1}{R(t)}(2\bar{P}_q-1)\right) \; ,\nonumber
\end{equation}

where $D_q$ is the pool-level angular diffusion coefficient (subscript $q$ signifying the pool level), and the previous noise-related diffusion coefficient $D_n$ is contained in $R(t)$ (see Eq.~\eqref{eq:reducedLength}).

\begin{figure}
     \centering
     \includegraphics[width=\textwidth]{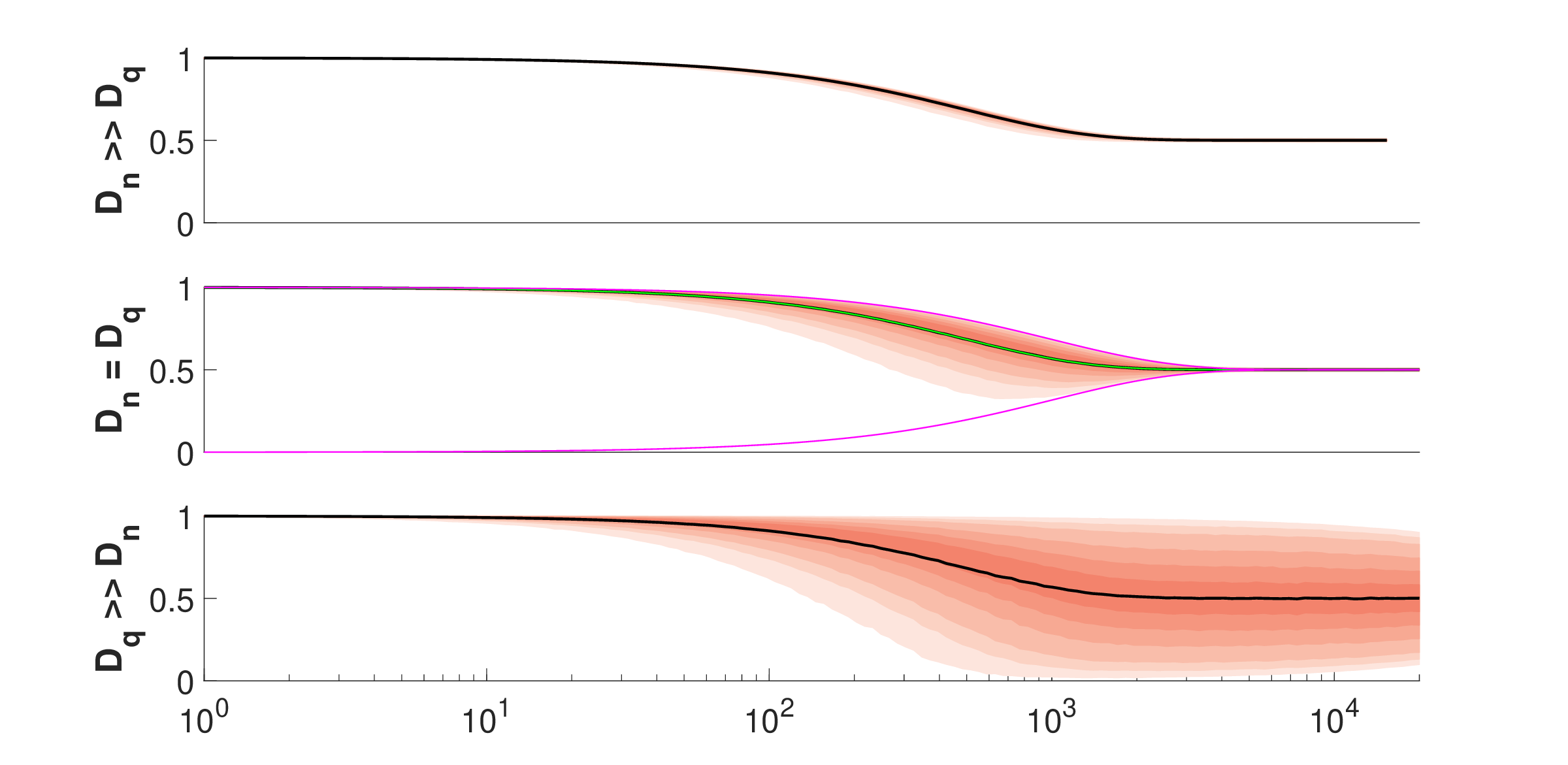}
     \caption{Simulated distribution of empirical frequencies over the number of applied gates for different relations between $D_n$ and $D_q$}
     \label{fig:Percentiles}
     \Description{For $D_n >> D_q$ the simulated curves only slightly disperse around the mean. For $D_n = D_q$ the simulated curves disperse more strongly around the mean while going towards the equilibrium, but not anymore once they arrive there. The distribution is nicely encapsulated by the analytically derived upper and lower limits. For $D_n << D_q$ the dispersion is largest and still visible after having reached the equilibrium.}
\end{figure}

% Relevant Properties
Although the combined distribution from Eq.~\eqref{eq:pPPtreducedLength} cannot be simplified analytically, we know several properties. To visualize them, Figure~\ref{fig:Percentiles} shows how the distribution of empirical frequencies $\bar{F}$ changes over ''time", i.e. over the number of applied gates, from $10^0$ to $2\cdot10^4$ on a logarithmic scale. The top plot shows a case of our model with dominant noise ($D_n \gg D_q$); the center plot shows the same for $D_n\approx D_q$, and the bottom plot shows the case of dominant pool-level errors ($D_q\gg D_n$). The relevant properties are:

\begin{itemize}
    \item \textbf{Upper bound:} At any given time $t$, the maximal attainable Binomial probability $\bar{P}$ occurs when $\theta_q=0$, and the corresponding probability $\bar{P}_{q,\text{max}}(t)$ is given by Eq.~\eqref{eq:pPPt}, and hence converges to 0.5 with exponential rate of $D_n$, see upper magenta line in the centre plot.
    
    \item \textbf{Lower bound:} At any given time $t$, the minimal attainable probability $\bar{P}$ occurs at $\theta_q=\pi$, and the corresponding probability is $\bar{P}_{q,\text{min}}(t) = 1- \bar{P}_{q,\text{max}}(t)$, see lower magenta line in the centre plot.
    
    \item \textbf{Total expectation:} The increments of the two random walks in angular coordinates are simply additive. Thus, the overall statistical expectation (i.e. the across-pool mean $\check{P}$) experiences a combined random walk with $D_{\text{tot}} = D_n+D_q$. Therefore, the across-pool mean $\check{P}$ has to evolve according to Eq.~\eqref{eq:pPPt}, with $D_{\text{tot}}$ substituted for $D_n$, see black line in all three plots. The black line always stays between (but not in the middle) of the two bounds. As we chose $D_q+D_n = const.$ the black line is always the same.
    
    \item \textbf{Overdispersion:} The overdispersion of Binomial-level probabilities $\bar{P}_q$ against the across-pool mean $\check{P}$ expressed as variance is given by Eq.~\eqref{eq:overdispersion} with diffusion coefficient $D_q$. The corresponding overdispersion is indicated in the plots via red-shaded percentile intervals. The corresponding overdispersion is hardly visible in the top plot, well visible in the center, and dominant in the bottom plot. 
       
\end{itemize}

% Wrap-up and overdispersion PDF shape
This means, we know time-dependent bounds for $\bar{P}_q(t)$, and we know its theoretical mean $\check{P}(t)$ and its theoretical variance as a function of $t$. We also know that the upper and lower bounds both converge to the same limit:
\begin{equation}
 \lim_{t\rightarrow\infty} \bar{P}_{q,  \text{max}}(t) =  \lim_{t\rightarrow\infty} \bar{P}_{q,\text{min}}(t) = \frac{1}{2}\, ,
\end{equation}

and that the limit distribution within these bounds for $t\rightarrow \infty$ is the $Sine()$ distribution. While the large-time limit ($Sine()$) distribution is apparently symmetric, the early-time distributions are concentrated at the upper bound. Accordingly, in the center plot, we can see how overdispersion initially does not fill the possible range between the limits, then opens up, and is finally suppressed by actual noise. In the bottom plot, we can see how the distribution of overdispersion asymptotically becomes symmetric, following a $Sine()$ distribution, before it would become suppressed by actual noise at time scales beyond the plotted range.

% ===========================================================+
\subsection{Bayesian inference of parameters in the probabilistic noise model\label{ssec:BayesInf}}
% ===========================================================+

% Notation for M and y
Having stochastically modeled the uncertainties at the Bernoulli, binomial, and pool levels, we set out to fit the model to our experimental data. We lump the diffusion coefficients on the different levels $D_n, D_q$ plus an initial $D_{ini}$ representing the SPAM noise into a model parameter vector $M$. Also, we arrange our experimental frequencies $\bar{F}$ after different lengths of quantum algorithms (see Section~\ref{subsec: data}) into a data vector $y$.

% clarification of data handling
We interpret one read-out frequency, i.e. a batch of 8192 algorithm executions with a certain number of gates, as a Binomial collection that undergoes a joint second-level random walk, meaning we look at all of the data in the same way as it is also presented in Figure~\ref{fig:singlerunspread}. We interpret every individual gate operation as a time step in both levels of the random walk and additionally let the Binomial-level random walk have a virtual zeroth time step with $D_{ini}$ as its diffusion coefficient.

% Bayesian analysis
As we are also interested in statements on significance and confidence about our model and its parameters $M$, we apply Bayesian parameter inference \cite{box2011bayesian}. That means, we seek the distribution $p\left(M|y\right)$ of model parameters $M$ conditional on data $y$:

\begin{equation}\label{eq:Bayes}
    p\left(M|y\right) \propto
    \int p\left(y|M\right)
        p\left(M\right)
    \text{d}\theta_q \, .
\end{equation}

Here, $p(M)$ is a prior PDF over the model parameters, set to be improper flat over the positive real numbers; $p\left(y|M\right)$ serves as the so-called likelihood function, expressing for any values in $M$ the model's quality of fit to the data $y$. In our case, it is given by the Binomial distribution, but only if we knew the values of $\theta_q$ within each Binomial ($n=8192$) trial, i.e. per observed frequency $\bar{F}$. Therefore, we treat the collection of $\theta_q$, one per Binomial-level trial, as hidden variables also to be inferred by the data, and use Eq.~\eqref{eq:thetaRQonSphere} as its PDF for any given value of $D_q$ substituted for $D_n$. 

% MCMC details
Technically, we use a Metropolis-Hastings algorithm \cite{chib2001markov} to solve Eq.~\eqref{eq:Bayes}, equipped with a so-called Gibbs split, where the proposals for the hidden variables $\theta_q$ and the primary parameters $M$ are done separately, and mutual independence among the entries of $\theta_q$ can be exploited. We generate $10^6$ parameter samples after burn-in from Eq.~\eqref{eq:Bayes} with our MCMC, out of which we store every 20th, and so obtain 50.000 approximately independent samples of $M$ to work with.

% ===========================================================+
\section{Results and discussion\label{sec:ResultsDiscussion}}
% ===========================================================+

% Posterior mean parameters
We apply our two-level model from Section \ref{subsec: TwoLevelRW} to the data from Section \ref{subsec: data} via Bayesian inference as described in Section \ref{ssec:BayesInf}. As mean conditional parameter values, we obtain $\Bar{D}_{ini} = 0.0218, \Bar{D}_n = 4.9764\times 10^{-4}, \Bar{D}_q = 3.2418\times 10^{-4}$, i.e. $D_n$ and $D_q$ are on a similar magnitude like in the center plot of Figure~\ref{fig:Percentiles}, and $D_{ini}$ for state preparation and measurement is about $50$ times larger, i.e. it is comparable to about $50$ initial $\sqrt{x}$ gate operations. 

\begin{figure}
     \centering
     \includegraphics[width=0.5\textwidth]{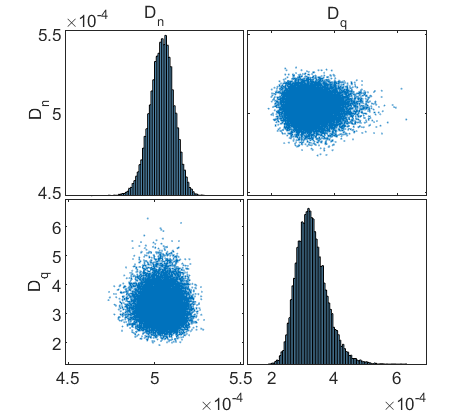}
     \caption{The (joint) distributions of $D_n$ and $D_q$}
     \label{fig:confusionMatrix}
     \Description{The plot shows the confusion matrix of the joint posterior of $D_n$ and $D_q$. $D_q$ is slightly skewed to the right. There is also a slight non-linear correlation visible in the scatter plot, with the skewness in $D_q$ mostly being derived from values at the center of the symmetric distribution of $D_n$.}
\end{figure}

% Posterior distributions and confirmation!
Figure~\ref{fig:confusionMatrix} shows the conditional distributions of $D_n$ and $D_q$ together with their joint distribution. Looking at the small scatter of inferred values relative to the inferred mean values, it becomes clear that both diffusion coefficients, including the one on the pool level, are distinctly different from zero to achieve a good model fit. In $10^6$ evaluated equiprobable cases, not a single realization shows $D_q$ anywhere below $2\cdot 10^{-4}$. This shows a substantial difference from zero for the pool-level $D_q$ with extreme confidence. Also, looking at the joint distribution, one cannot see any correlation between $D_n$ and $D_q$, i.e. more noise via larger $D_n$ could not compensate to replace $D_q$.  The claim that the pool level does indeed improve the model is strengthened further by comparing the maximal found log-likelihood values to the single-level model (obtained by simply forcing $D_q=0$). Here, we have values of $\mathcal{L}(M_1) = -8974$ and $\mathcal{L}(M_2) = -1186$, a log-ratio of about 7000 in favor of our two-level model. 

\begin{figure}
     \centering
     \includegraphics[width=\textwidth]{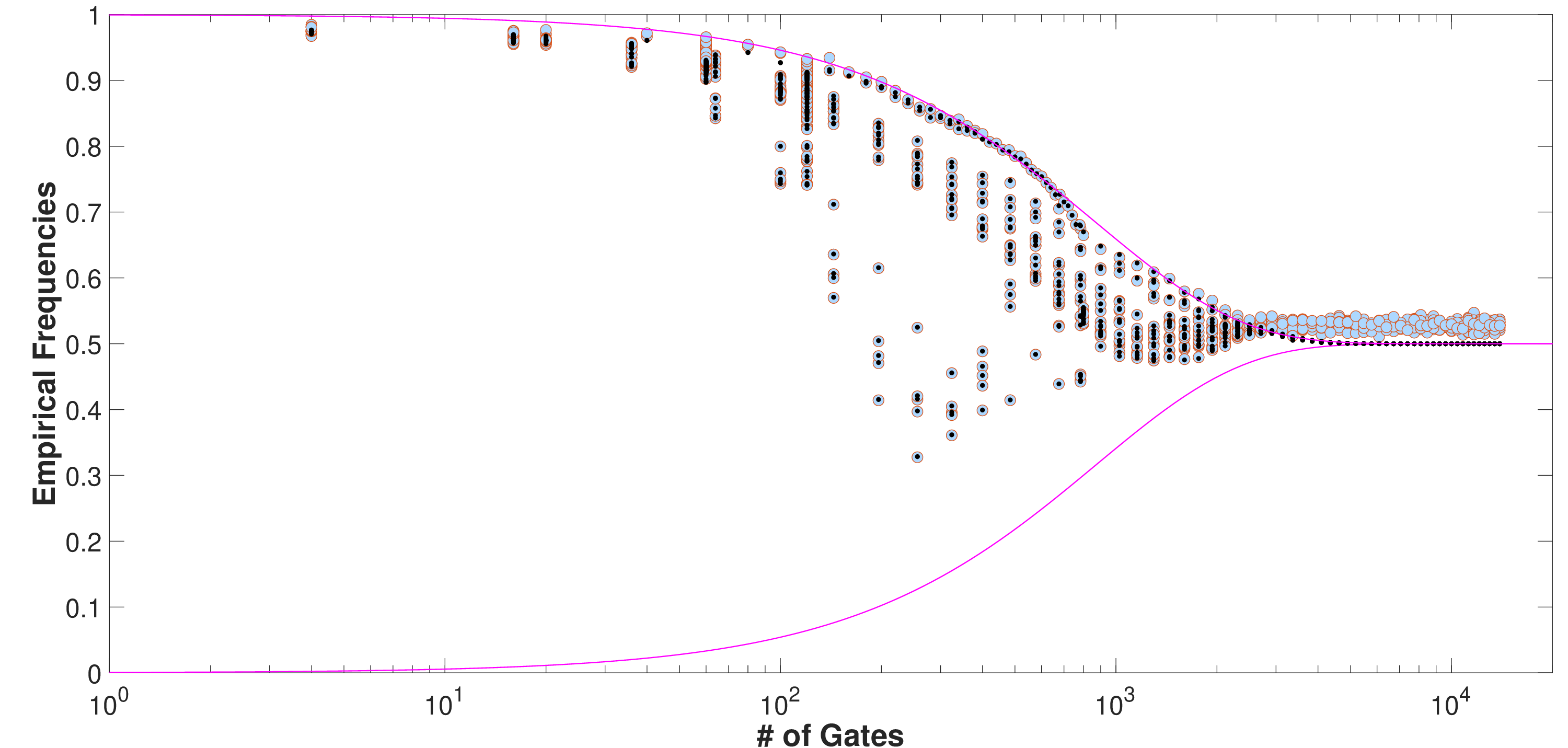}
     \caption{How well the empirical frequencies are matched}
     \label{fig:P_match}
     \Description{The analytically derived limits for the MAP parameters yielded by the inference nicely encapsulate the experimental results in all parts but the equilibrium, which experimentally lands a bit above the expected value of $0.5$. The hidden variables of the inference also nicely match the actual distributions at each number of gate applications.}
\end{figure}

% Model fits super well
Figure~\ref{fig:P_match} shows the resulting fit of the model to the data. Specifically, the inferred probabilities $\bar{P}_q$ at the Binomial level, based on the hidden variables $\theta_q$ (black dots), nicely match the observed frequencies $\bar{F}_q$ (blue circles). Also, the data neatly falls in between the two theoretical bounds predicted by our model and populates especially the space close to the upper bound. Therefore, $D_n$ (the noise-related diffusion coefficient that dictates the upper bound) could almost be found directly by graphical analysis. As additional visual evidence in favor of our two-level model, one can see that trying to fit the data from Figure~\ref{fig:P_match} into the upper plot of Figure~\ref{fig:Percentiles} is quickly rejected, i.e. the pool-level random walk with its resulting explanation for overdispersion is indeed required.

% discrepancy at late times
There remains a noticeable discrepancy between the data and the model in the late stages. However, this long-time behavior is not of interest when it comes to characterizing the error for quantum algorithms, as the distribution is almost completely uninformative by then anyway. In ad-hoc modeling variants not shown here, we improved this aspect by introducing a different asymptote, with another parameter to be tuned, into the model, giving even better visual fit and likelihood values.

\begin{figure}
     \centering
     \includegraphics[width=\textwidth]{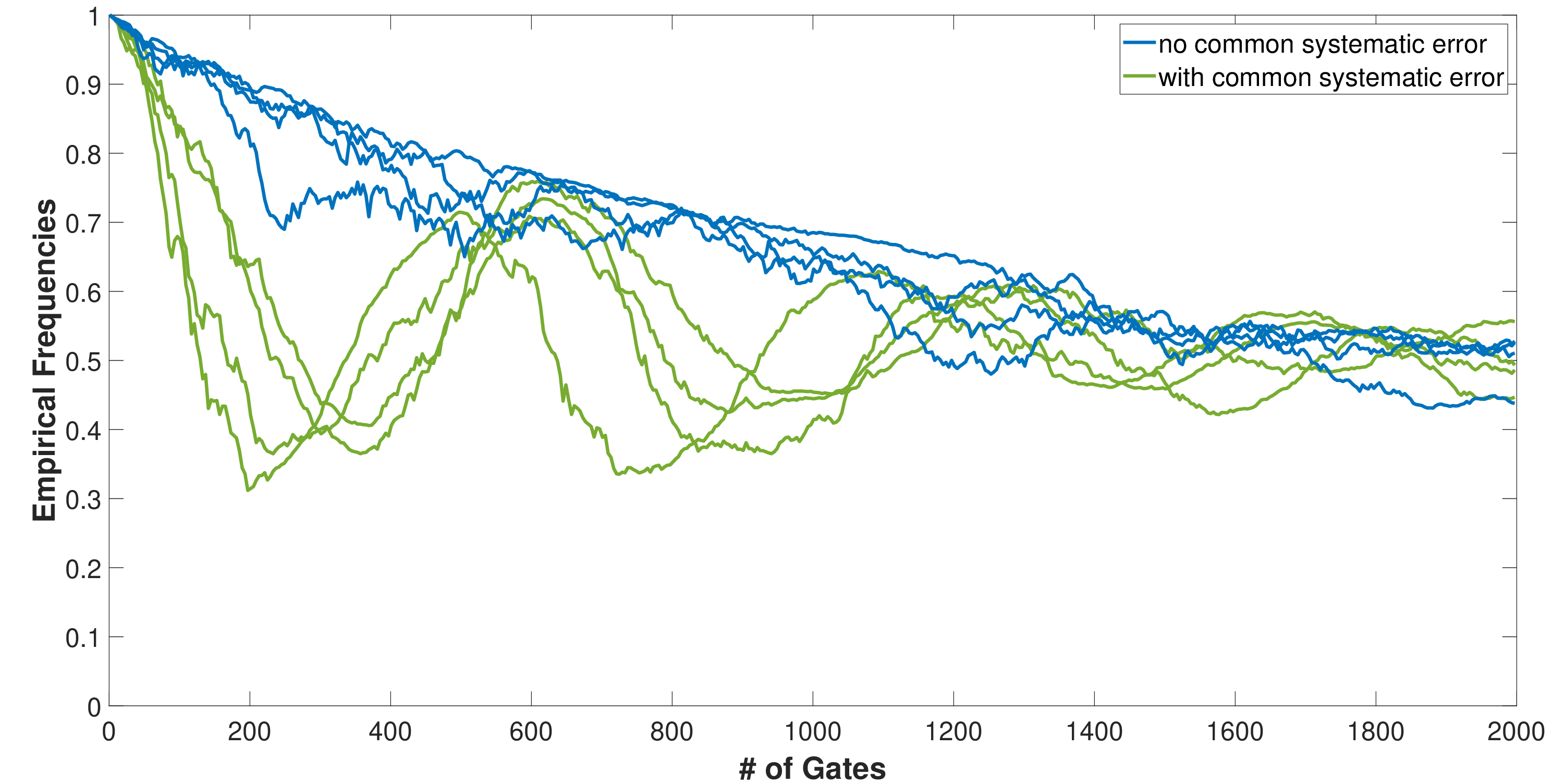}
      \caption{Forward runs simulated with the MAP estimates. The green curves have an added common deterministic error per run, resulting in oscillations comparable to Figure \ref{fig:multirun}.}
     \label{fig:resampling}
     \Description{Plot showing that adding a simplistic coherent error to the simulations makes the resulting resampled curves visually match the behavior found for single queue experiments.}
\end{figure}

% Oscillations missing!
As a last check, we perform Monte-Carlo simulations of the random walk to see how closely we could match the optics of a figure like Figure~\ref{fig:multirun}. Results are shown in Figure~\ref{fig:resampling}. Many of the runs in Figure~\ref{fig:multirun} display oscillations, with systematic errors as the most likely explanation. These oscillations occur at some times of the day, and not at others, so some curves show this behavior, while others do not. Our model reproduces the time-sliced data in the distributional sense. The blue curves in Figure~\ref{fig:resampling} represent repeated gate applications simulated with the parameters obtained from the distributional, time-sliced viewpoint. To emulate the behavior of the oscillatory runs, a shared systematic over-rotation was added to a constant fraction of the qubits producing the green curves in Figure~\ref{fig:resampling}. The specific quantities stem from manual ad-hoc tuning, but it is possible to also include such parameters in the Bayesian inference if one has enough data to use full runs as data points. As we set out to characterize random (diffusive) rather than coherent errors as a function of algorithmic lengths, we omit this aspect in our current analysis.

% possible extension: non-markovian
Nevertheless, a possible extension of our model would be to add to the current pool-level random walk a coherent component, which has increments that are constant or autocorrelated over wallclock time. In the diffusion literature, such non-Markovian behavior is well known. Then, the diffusion coefficient, constant in our model, is defined as half the rate of growth of the variance:

\begin{equation*}
      D_x(t) = \frac{1}{2} \frac{\partial \sigma^2_{x}}{\partial t} \, .
\end{equation*}

Thus, we could modify our above results by re-scaling the time coordinate with an exponent $0\leq \kappa \leq 2$ to obtain

\begin{equation*}
    \bar{P}(t) = \frac{1}{2} + \frac{1}{2} \exp\left(-2 D_{r}  \Delta t \left(\frac{t}{\Delta t}\right)^{1+\kappa}\right) \, .
\end{equation*}

This could be done separately at the pool level and the Binomial levels, introducing two new parameters into our model to alleviate the assumption of Markovianity behind Eq.~\ref{eq:RandomWalkOnSphere}.

% ===========================================================+
% ===========================================================+
\section{Conclusion and outlook}
\label{sec:conclusion}
% ===========================================================+
% ===========================================================+

% summary of model idea
In this study, we investigated the issue of overdispersion in noise that appears on NISQ machines. We proposed a two-level random walk model on the Bloch sphere and tested it against experimental data obtained from the IBM Armonk NISQ machine. Our model assumes that the effect of classical noise can be represented as individual random walkers on the Bloch sphere, while overdispersion is a pool-level random walk that jointly affects the theoretical center of mass of all algorithmic repetitions that are used in quantum computing to obtain observable read-out frequencies. For given parameters (diffusion coefficients for noise, for overdispersion, and for state preparation and measurement), it predicts the probability distribution of observable, noisy frequencies as a function of number of quantum gates applied. It also predicts theoretical bounds for the observable frequencies. 

% success of model quality
Our model fits very well with the experimental data. We employed Bayesian calibration to highlight the necessity of the newly introduced pool-level. It is substantial in strength and statistically significant in its effect, with a log-likelihood ratio of more than 7000 in favor of our new model, compared to a conventional single-level random walk model. In practice, computationally cheaper optimization algorithms can be used for calibrating the model, e.g. by maximizing Eq.~\eqref{eq:Bayes} with respect to the noise-model parameters, if the parametric uncertainty is not of interest.

% remaining limitations and outlook
Several limitations and  aspects for future investigation became apparent:

\begin{itemize}
    \item The late-time limit for long algorithms in our experimental data is clearly offset against the theoretical 50:50 probabilities. It would be interesting to apply models about the equilibrium (energy-optimal) distribution of qubit states to replace the 50:50 asymptote in our model with an offset asymptote.
    
    \item Parts of the experimental data show oscillatory behavior along data acquisition, pointing at coherent error types. Our model could be extended to relax its underlying Markovian assumption in its random walks to account for coherent errors.
    
    \item The omission of the non probability relevant angle $\varphi$ enables the semi-analytic description of the error model presented. While diffusive errors in this angle can be incorporated by the parameters of the model, an explicit modelling of this additional angle would enable representations of many more relevant effects due to systematic errors.
    
    \item Our study's scope was focused on data from a single machine (IBM Armonk), which is a single-qubit architecture. This is a limited study based on specific empirical observations. It would definitely be interesting to test the two-level approach on other hardware types. Even though highly speculative, we would anticipate some generalization properties. The two-level structure is partially motivated by two apparent error pathways, inherent per qubit randomness and shared randomness due to environmental effects. This initial motivation still holds for other hardware types.
\end{itemize}

\begin{acks}
We thank the \grantsponsor{dfg}{Deutsche Forschungsgemeinschaft (DFG, German Research Foundation)}{} for supporting this work by funding - \grantnum{dfg}{EXC2075 – 390740016} under Germany's Excellence Strategy. We acknowledge the support by the Stuttgart Center for Simulation Science (SimTech). We furthermore thank the \grantsponsor{diam}{Delft Institute of Applied Mathematics (DIAM)} for the financial support of Merel Schalkers and acknowledge the use of IBM Quantum services for this work. Moreover, we thank the \grantsponsor{sfb}{Collaborative Research Centre SFB 1313, Project Number 327154368}{} for the support of Tim Brünnette.
\end{acks}

\appendix

\section{Shifted Legendre polynomials} \ignorespacesafterend
For completeness, we provide here the first three shifted Legendre polynomials:
\begin{eqnarray}
\mathcal{L}_0(X) &=& 1 \\
\mathcal{L}_1(X) &=& 2x-1\\
\mathcal{L}_2(X) &=& 6x^2-6x+1
\end{eqnarray}

\section{Qiskit specifications} \label{app:ArmonkSettings}
We run our experiments on the open-access 1-qubit IBM Armonk machine \cite{IBM}. The circuits were created and run using the following versions the qiskit packages {'qiskit-terra': '0.19.2', 'qiskit-aer': '0.10.3', 'qiskit-ignis': '0.7.0', 'qiskit-ibmq-provider': '0.18.3'}. These packages are all part of the qiskit framework and provide the user with the needed functionality to write and perform quantum circuits on the IBM hardware or simulators. The exact explanation of the functions contained in each package and version can be found on the IBM quantum website \cite{IBM}.

\bibliographystyle{ACM-Reference-Format}
\bibliography{lit}

\end{document}